    \documentclass[11pt,a4paper]{article} 
    \usepackage{graphics,subfigure}

\def\Journal#1#2#3#4{{#1}{\bf #2}(#4) #3}

\def\PLB{{ Phys.\ Lett.}  \bf B}

\def\ZPC{{ Z.\ Phys.} \bf C}

\def\lsim{\mathrel{\rlap{\lower4pt\hbox{\hskip1pt$\sim$}}
    \raise1pt\hbox{$<$}}}                
\def\gsim{\mathrel{\rlap{\lower4pt\hbox{\hskip1pt$\sim$}}
    \raise1pt\hbox{$>$}}}                

\begin{document}

\vspace{1.0cm}
\vspace*{1.0cm}
\begin{center}  \begin{Large} \begin{bf}
Colour Reconnection at LEP \footnote{Proceeding of the EPS conference on HEP, Aachen 2003} \\
  \end{bf}  \end{Large}
  \bigskip
  \bigskip

\vspace{1.0cm}
  \begin{large}
Jorgen D'Hondt \\
  \end{large}
\medskip

  {\it Inter-university Institute for High Energies} \\
{\it Vrije Universiteit Brussel} \\
{\it Pleinlaan 2, B-1040 Brussels, Belgium} \\
{\it email : jdhondt@hep.iihe.ac.be} \\
\end{center}
\bigskip
\vspace{1.5cm}
\abstract{
Two measurements are presented of estimators sensitive to the 
Colour Reconnection effect in $\mathrm{W^+W^-}$ events at LEP2. The results are
compared with various phenomenological Monte Carlo implementations 
of the effect. A feasibility study is performed to reduce
the total uncertainty in the direct $\mathrm{m_W}$ measurement at LEP2 by use of 
the inferred information about the Colour Reconnection effect.
}%
%
%
\section{Introduction}
\label{intro}

Colour Reconnection is the term used for strong interactions between colour singlet 
parton systems of different origin~\cite{gosta1}. The effect
can influence the evolution of two nearby parton showers.
The kinematics of the hadrons coming from both systems can therefore
be perturbed. In the reaction $\mathrm{e^{+}e^{-} \rightarrow W^{+}W^{-} \rightarrow 
q_1 \bar{q}_2 q_3 \bar{q}_4}$ the colour singlets $\mathrm{q_1 \bar{q}_2}$ and
$\mathrm{q_3 \bar{q}_4}$ formed by the decay products of both W bosons, could
rearrange themselves to new colour singlets $\mathrm{q_1' \bar{q}_4'}$ and
$\mathrm{\bar{q}_2' q_3'}$. Because the flow of the colour quantum numbers, reflecting the particle
dynamics at short distances, controls the particle distributions in the final state, one
could expect a change in these distributions after the colour rearrangement.
In these $\mathrm{W^+W^-}$ events produced at LEP2, the separation distance between the $\mathrm{W^{\pm}}$ decay
vertices is around $\mathrm{\tau_W \sim 1/\Gamma_W \simeq 0.1}$ fm, while the fragmentation scale of the W bosons is around 1 fm.
Hence there is a significant space-time overlap of both W hadronization regions where Colour
Reconnection could occur. The resulting kinematic structure of a $\mathrm{W^+W^-}$ event will be different from
the situation without this reconnection.
The presence of similar effects was found in hadronic B meson decays $\mathrm{B \rightarrow J/\psi + X}$ where the colour
interference between the two original colour singlets ($\mathrm{{\bar c}+s}$ and c+spectator) suppresses this decay~\cite{gosta1}.

The probability to rearrange can be enhanced by gluon
exchange between both W decay systems. Within the perturbative parton shower model
it was shown that the Colour Reconnection or interference probability is 
negligible~\cite{sk}. Colour transmutations between partons from different W bosons can
only occur from the interference of at least two emitted gluons. This interference
piece is suppressed by the effective number of colours as $\mathrm{1/N_c^2 = 1/9}$ compared to the $\mathrm{{\cal O}(\alpha_s^2)}$ 
non-reconnection emissions. Also the effects of a finite W width restrict the energy range
of primary gluons generated by the alternative rearranged dipoles, because both W bosons
do not necessarily decay at the same time.

Within the framework of the Lund string fragmentation implemented in JETSET the 
colour fields of both W boson strings can overlap in space-time.
When the strings are described in their simplest Lorentz invariant
way by a linear confinement potential, the event probability $\mathrm{{\cal{P}}_i}$ 
to reconnect is related to the volume, $\mathrm{{\cal{O}}_i}$, of the string overlap as:

\begin{equation}
\mathrm{ {\cal{P}}_{i} = 1 - e^{-\kappa \cdot {\cal{O}}_i}} \ \ .
\label{eq:sk1}
\end{equation}

\noindent
The SK1 model parameter $\mathrm{\kappa}$, with a fixed value for each event, is unknown and 
can only be tuned or measured from experimental data. For simplicity only 
one reconnection per event was allowed.

Within ARIADNE the Dipole Cascade Model is also followed by a string fragmentation
according to the Lund model.
As a criterion for Colour Reconnection 
one can therefore use the string length $\mathrm{\Lambda}$ defined in the momentum space 

\begin{equation}
\mathrm{\Lambda = \sum^{n-1}_{i=1} \frac{ln(p_i + p_{i+1})^2}{m_0^2}}
\end{equation}

\noindent
where the sum covers all $\mathrm{n-1}$ string pieces for a string connecting $\mathrm{n}$ partons
and $\mathrm{m_0}$ is a typical hadronic mass scale (usually $\mathrm{\Gamma_W}$). Reconnection 
is allowed when the string length after
reconnection is shorter than the original one. In the ARIADNE 2 (AR2) model, 
reconnection between strings from different W bosons only happens below a fixed
energy scale (usually $\mathrm{\Gamma_W}$), while in the ARIADNE 3 (AR3) version 
reconnection is allowed at all energy scales which includes the perturbative phase.
Therefore it is only the AR2 model which is relevant for this study.
Within an event multiple inter-string reconnections and self-reconnections inside 
a single string could occur.

\begin{figure*}
\begin{center}
\hspace{-0.8cm}
\subfigure[]
{ \resizebox{0.49\textwidth}{!}{\includegraphics{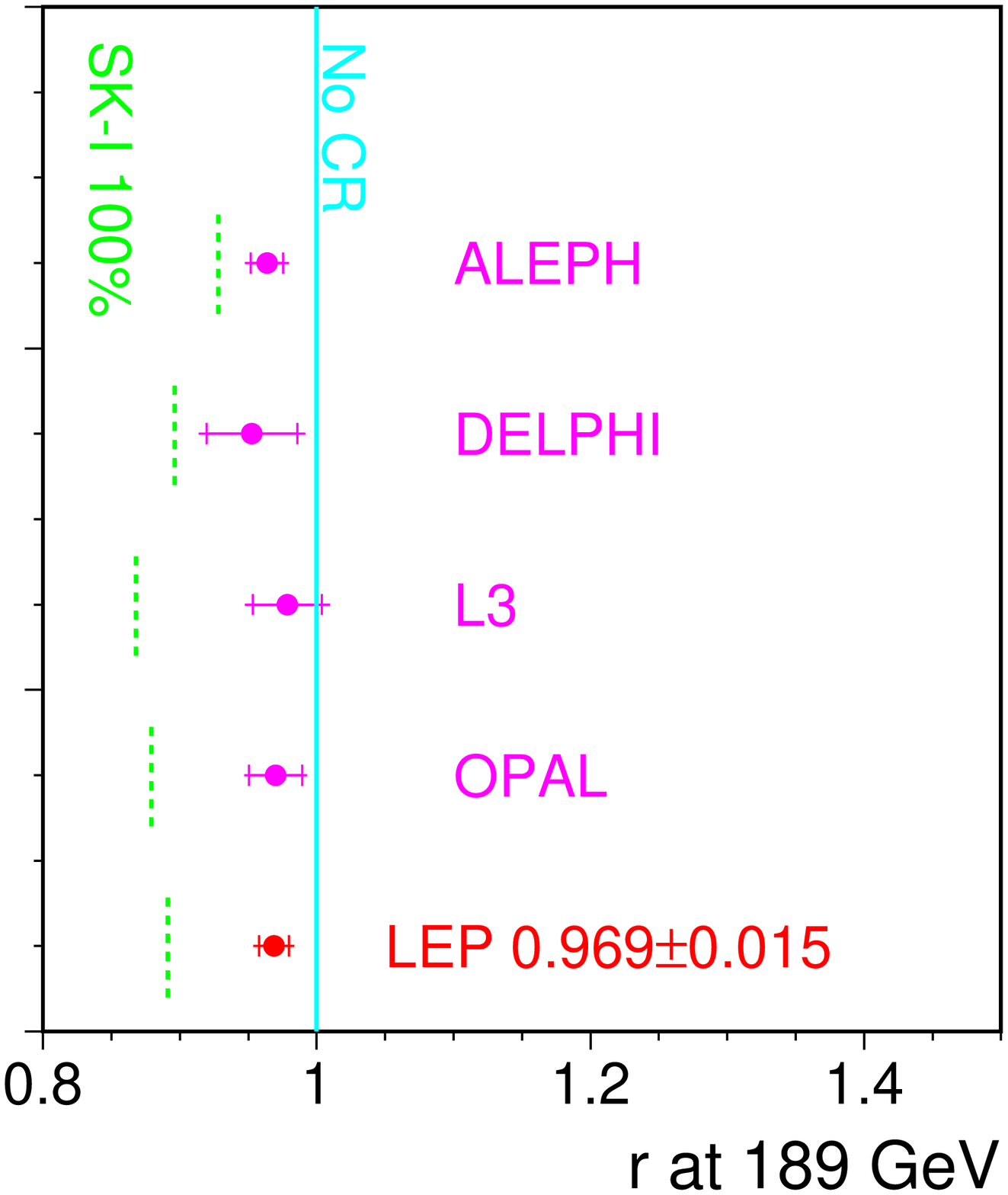}}}
\subfigure[]
{ \resizebox{0.49\textwidth}{!}{\includegraphics{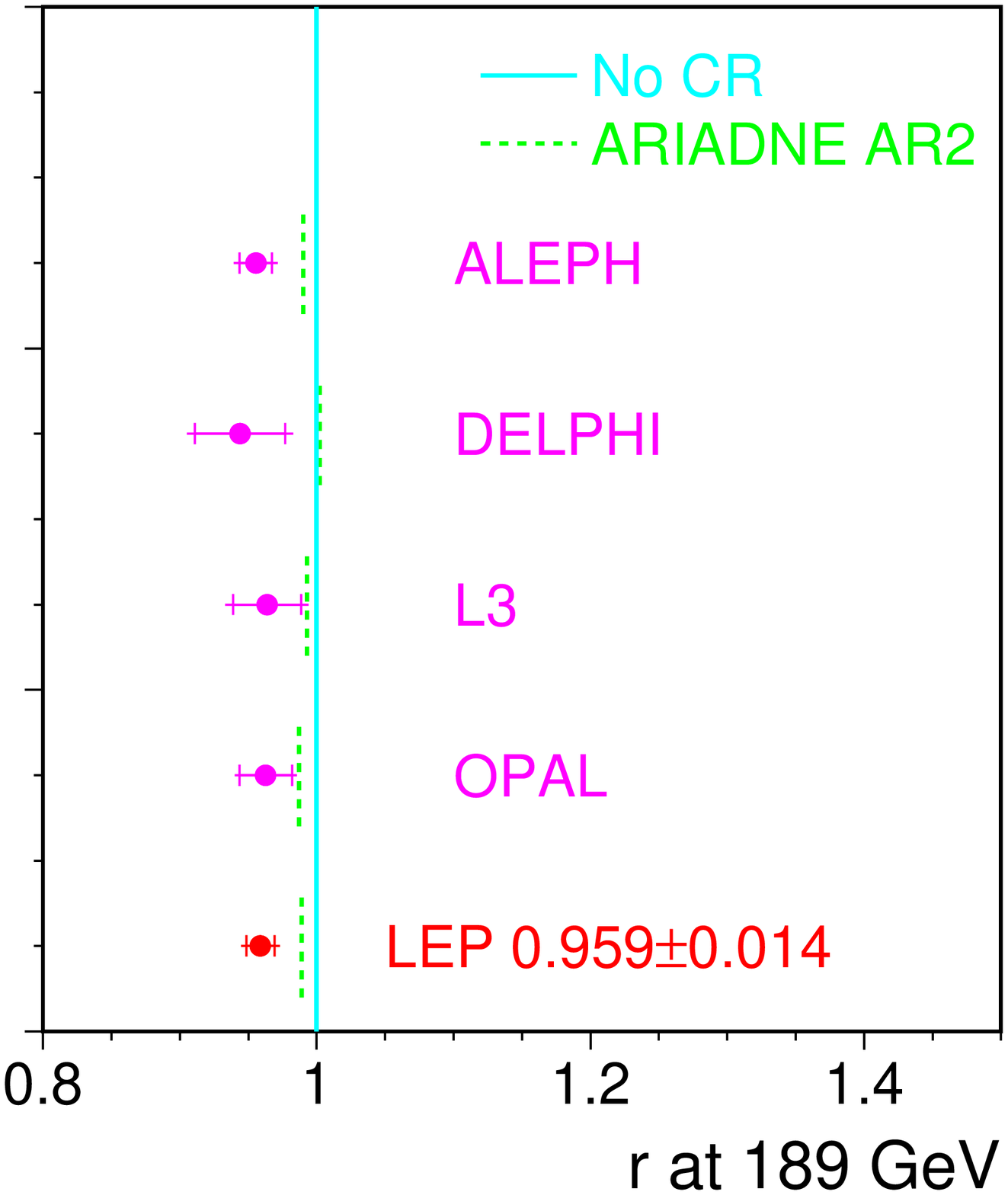}}}
\subfigure[]
{ \resizebox{0.49\textwidth}{!}{\includegraphics{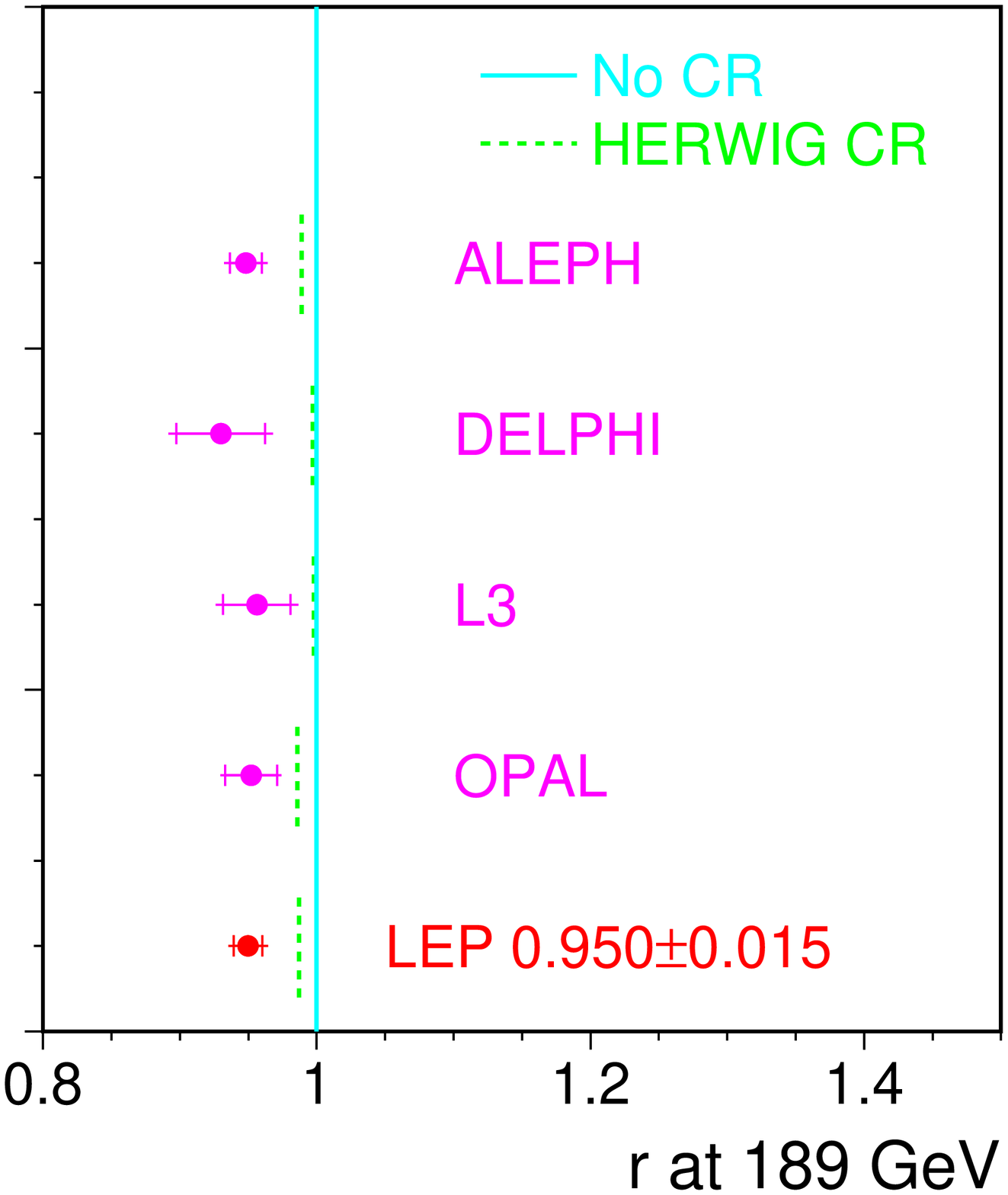}}}
  \caption{Preliminary particle flow results on $\mathrm{r=R^{data}/R^{no-CR}}$ using all LEP2 data. 
In plot (a) both hypotheses without and with full (100\%) SK1
Colour Reconnection are shown, in plot (b) and (c) the ARIADNE and HERWIG Colour Reconnection models are tested,
based on the predicted sensitivity. The dashed lines indicate the predicted values of $\mathrm{r}=R^{CR}/R^{no-CR}$ 
for the analysis of each 
experiment. For comparison all values are interpolated to a centre-of-mass energy of 189 GeV and the LEP combined values
are shown on the bottom of each plot.}
\end{center}
\label{fig:particleflow}
\end{figure*}

Before the formation of the clusters in the HERWIG fragmentation process, reconnection is
allowed with a fixed probability ($\mathrm{\simeq}$ 1/9) when it reduces the sum of the squared sizes 
of the formed clusters. The size of a
cluster is defined as the space-time separation of the production vertices of the partons
forming the cluster. Multiple reconnections and self-reconnections within a single 
parton shower are allowed.


\section{Particle flow measurement}
\label{sec:particleflow}

The string effect predicts a higher particle production rate in the region between jets originating
from the same $\mathrm{W \rightarrow q{\bar q}}$ decay (`intra' region), compared to the regions between 
jets from different W decays (`inter' region). When Colour Reconnection is present, particles tend to
migrate from the intra-W to the inter-W region. Therefore an observable which counts the particles in these
different regions could be sensitive to the different Colour Reconnection algorithms. The ratio $\mathrm{R}$ of the 
integrated particle density or particle flow in the intra region to the integrated density in the 
inter region quantifies this behavior. 

The value obtained from the LEP combined data is compared to the
different models in Figure 1. In the SK1 interpretation of Colour Reconnection the data
prefers a value of $\mathrm{\kappa}$ = 1.18, and the 68\% confidence level lower and upper limits are 0.39 and 2.13 
respectively. This corresponds to a reconnection probability of 49\% in this model at $\mathrm{\sqrt{s}}$ = 189 GeV.
The extreme SK1 scenario with 100\% reconnection probability disagrees with the data at a level
of 5.2 $\mathrm{\sigma}$. The log-likelihood as a function of $\mathrm{\kappa}$ is shown in Figure~\ref{fig:comb}.
All four experiments have observed a very weak sensitivity to the ARIADNE and HERWIG Colour 
Reconnection models, which does not coincides with the SK1 prediction of the effect.
The expected value of $\mathrm{R}$ from those fragmentation models in the hypothesis that no Colour
Reconnection is present, differs from the measured LEP combined data value by 3.1 and 3.7 $\mathrm{\sigma}$ for respectively the
ARIADNE and HERWIG model. This may indicate an underestimation of the systematic uncertainties in
these particle flow analyses.
The dominating systematic uncertainties are the fragmentation modeling in the simulation 
and the possibility for Bose-Einstein Correlations in the final state.

\section{The $\mathrm{\bf \Delta m_W}$ method~\cite{dmw}}
\label{sec:dmw}

It has been shown that the $\mathrm{m_W}$ observable~\cite{ew} inferred from hadronic decaying 
$\mathrm{W^{+}W^{-}}$ events at LEP2 by the method of direct reconstruction, 
is influenced when changing the value of $\mathrm{\kappa}$. 
A second method is therefore based on the observation that two different $\mathrm{m_W}$ 
estimators have different sensitivity to the parametrized Colour 
Reconnection effect. Hence the difference between them is an 
observable with information content about $\mathrm{\kappa}$. 

It is observed that mostly low momentum particles and 
particles in inter-jet regions are affected by Colour Reconnection and hence influencing the 
measurant, $\mathrm{m_W}$. An alternative analysis was designed
which neglects these particles in the reconstruction of the momenta of the four
primary partons. This by applying an optimized cone algorithm to cluster the particles 
in jets, rather than using inclusive clustering algorithms like DURHAM.

An indirect measurement of the SK1 model parameter $\mathrm{\kappa}$ 
is possible from the direct measurement of the difference $\mathrm{\Delta m_W(i,j)}$ in 
reconstructed $\mathrm{m_W}$ between the CR-sensitive standard (i) and CR-less-sensitive 
alternative (j) analysis. 
The observable $\mathrm{\Delta m_W(std,R_{cone})}$ where $\mathrm{R_{cone}}$ is 
around 0.5 rad, was found to be most sensitive (4.3 $\mathrm{\sigma}$ sensitivity to the full SK1 model). 
The systematic uncertainty is dominated by possible effects like Bose-Einstein Correlations
or detector related discrepancies in the energy flow reconstruction.
The log-likelihood as a function of $\mathrm{\kappa}$ obtained
by DELPHI is shown in Figure~\ref{fig:comb}.

After rescaling the effect according to the overall reconnection probability difference between the
SK1 and the HERWIG model, also similar sensitivities were observed for both models.
Only a negligible sensitivity to the ARIADNE implementation was found.

The correlation between the $\mathrm{\Delta m_W(std,R_{cone}=0.5 rad)}$ and standard $\mathrm{m_W}$ estimators was
found to be around 11\%.

\section{Feasibility study for the $\mathrm{\bf m_W}$ measurement}
\label{sec:mw}

\begin{figure}
\vspace{-0.5cm}
\resizebox{0.95\textwidth}{!}{\includegraphics{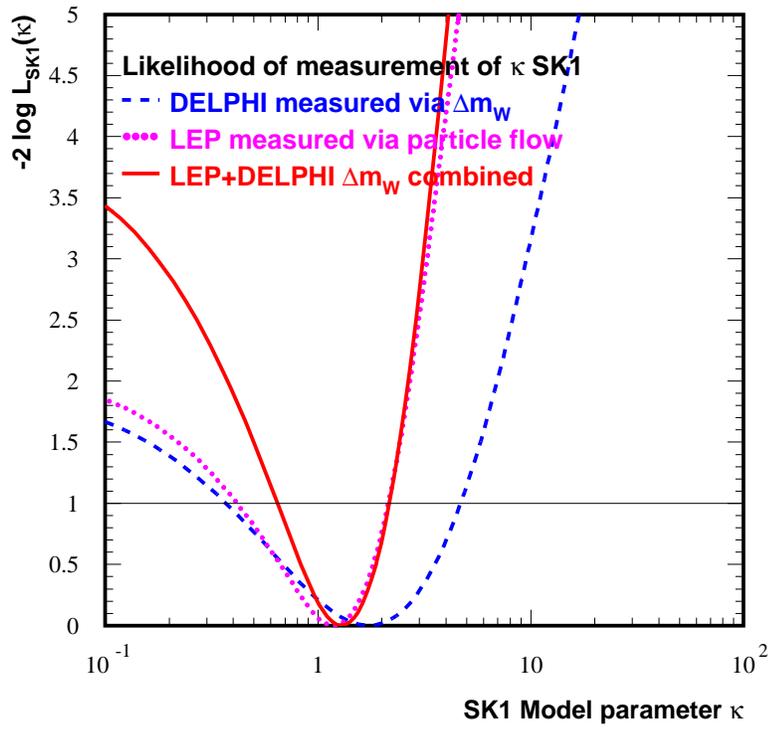}}
\vspace{-0.4cm}
\caption{Combined log-likelihood information on $\mathrm{\kappa}$ from both measurement 
assuming they are uncorrelated.}
\label{fig:comb}
\end{figure}

The log-likelihood information about $\mathrm{\kappa}$ from both measurements is
combined in Figure~\ref{fig:comb}. Within the SK1 model a significant amount of
Colour Reconnection is found.

\begin{figure}
\vspace{-0.5cm}
\resizebox{0.95\textwidth}{!}{\includegraphics{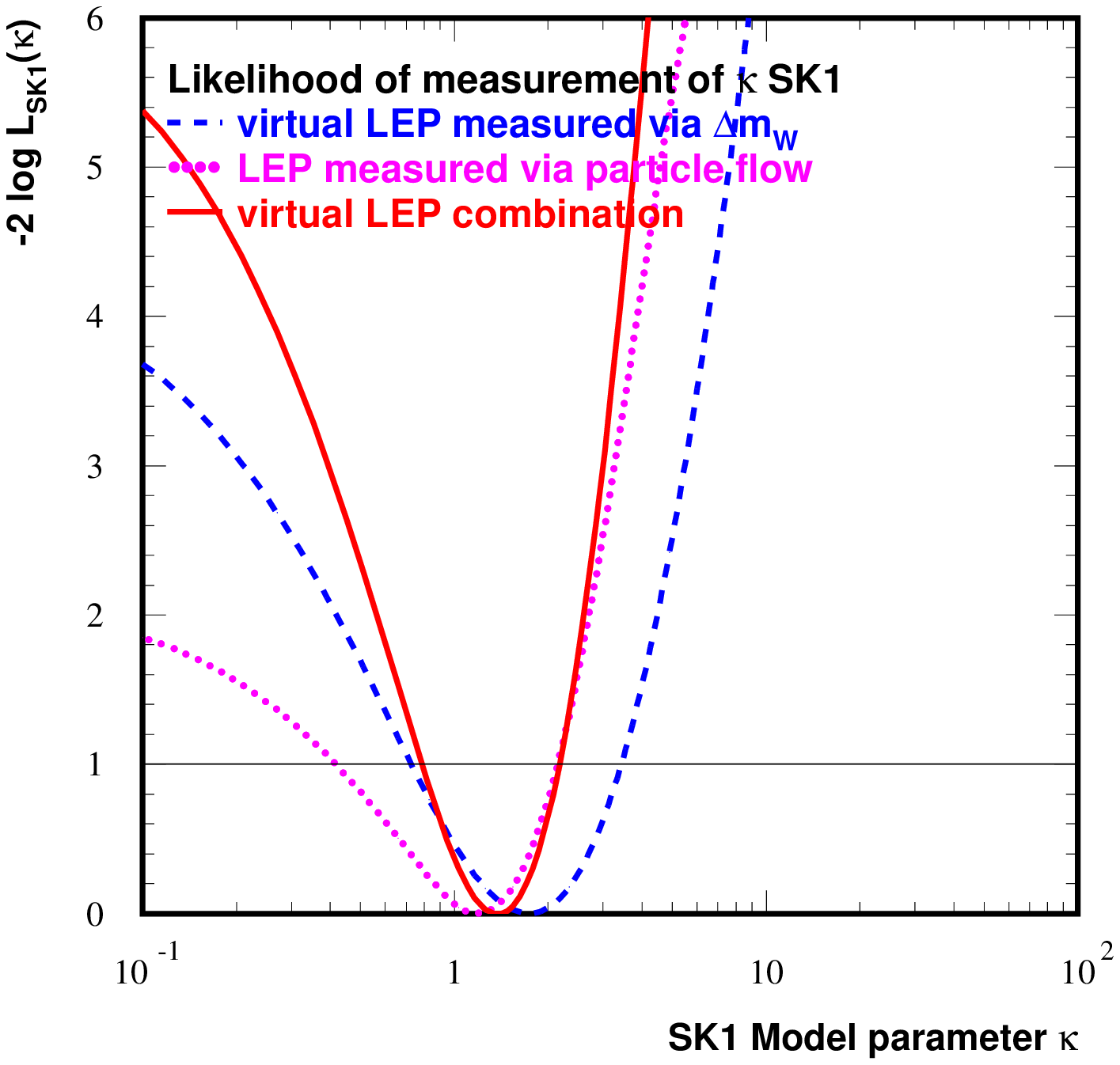}}
\vspace{-0.4cm}
\caption{Feasible log-likelihood information about Colour Reconnection to be extracted
by the LEP2 data in WW events. The DELPHI $\mathrm{\Delta m_W}$ log-likelihood is 
rescaled to four times more luminosity.}
\label{fig:feas}
\end{figure}

The $\mathrm{m_W}$ estimators applied by the four LEP Collaborations are equally
sensitive to the effect of Colour Reconnection. Therefore one can expect to
infer the same amount of likelihood information from all four datasets. In this 
hypothesis and assuming that the $\mathrm{\Delta m_W}$ analysis would reveal the same
most likely value for $\mathrm{\kappa}$, the total log-likelihood obtained by LEP2 would
be Figure~\ref{fig:feas}. 
For $\mathrm{m_W}$ measured in the fully hadronic channel 
the minimum of this log-likelihood corresponds to a shift
of 106 MeV/c$^2$ at $\mathrm{\sqrt{s}}$ = 200 GeV with a 68\% CL of 
[77;144] MeV/c$^2$ (where -2log$\mathrm{\cal L}$=1). 
These numbers should be compared with the statistical uncertainty of 28 MeV/c$^2$ in the
absence of systematic uncertainties~\cite{ew03}.
Assuming that the SK1 model
is the true model of Colour Reconnection, one could calibrate for this shift. The remaining
uncertainty due to Colour Reconnection would be about 35 MeV/c$^2$ (neglecting ARIADNE and HERWIG). 
This would reduce the total uncertainty on
$\mathrm{m_W}$ measured in the fully hadronic channel from about 110 to 60 MeV/c$^2$ and
increase the relative weight of this channel from 9 to 29\% in the combination with
the semi-leptonic channel.

\section{Conclusion}
\label{sec:concl}

Two basically uncorrelated measurements observe the same amount of Colour
Reconnection in the LEP2 data according to the SK1 phenomenological model.
Within the direct measurement of $\mathrm{m_W}$ this effect was not
corrected for and therefore induced a significant systematic uncertainty.
A simple but efficient calibration method was proposed to benefit from the inferred
information about the Colour Reconnection effect in the $\mathrm{m_W}$ measurement.
The uncertainty on the LEP2 value of the direct measurement of 
$\mathrm{m_W}$ would reduce from 42 to 39 MeV/c$^2$.


\end{document}